\documentclass[preprint,12pt]{elsarticle}

\usepackage{tabularx} 
\usepackage{epsfig}
\usepackage{bm}
\usepackage{graphicx}
\usepackage{float}
\usepackage{gensymb}
\usepackage{amsmath} 
\usepackage{amssymb} 
\usepackage{subfig}
\usepackage{tensor}
\usepackage[hidelinks]{hyperref}
\usepackage{wrapfig}
\usepackage{color}
\usepackage{xcolor}
\usepackage{caption}
\usepackage{multirow}
\usepackage{hhline}

\colorlet{darkgreen}{green!50!black}
\colorlet{brightyellow}{yellow!75!red}
\colorlet{orange}{red!50!yellow}
\colorlet{darkblue}{blue!60!black}
\colorlet{darkred}{red!80!black}

\newcommand{\comment}[1]{}

\newcommand{\pcal}{\mathcal{P}}

\usepackage[normalem]{ulem}
\renewcommand{\sout}{\bgroup \color{red} \ULdepth=-.5ex \ULset}

\DeclareCaptionLabelFormat{cont}{#1~#2\alph{ContinuedFloat}}
\journal{Annals of Physics}

\begin{document}

\begin{frontmatter}

\title{Angular Momentum Eigenstates of the Isotropic 3-D Harmonic Oscillator: Phase-Space Distributions and Coalescence Probabilities}

\author{Michael Kordell II}

\author{Rainer J.\ Fries}
\ead{rjfries@comp.tamu.edu}

\author{Che Ming Ko}

\affiliation{organization={Cyclotron Institute and Department of Physics and Astronomy, Texas A\&M University},
   city={College Station, TX}, postcode={77845}, country={USA}}



\begin{abstract}
The isotropic 3-dimensional harmonic oscillator potential can serve as an approximate description of many systems in atomic, solid state, nuclear, and particle physics. In particular, the question of 2 particles binding (or coalescing) into angular momentum eigenstates in such a potential has interesting applications. We compute the probabilities for coalescence of two distinguishable, non-relativistic particles into such a bound state, where the initial particles are represented by generic wave packets of given average positions and momenta. We use a phase-space formulation and hence need the Wigner distribution functions of angular momentum eigenstates in isotropic 3-dimensional harmonic oscillators. These distribution functions have been discussed in the literature before but we utilize an alternative approach to obtain these functions. Along the way, we derive a general formula that expands angular momentum eigenstates in terms of products of 1-dimensional harmonic oscillator eigenstates.
\end{abstract}

\end{frontmatter}


\section{Introduction}

Harmonic oscillator potentials are commonly used to approximate the low-energy behavior of attractive interactions between two particles. Their applications range from vibrational states in molecules to the excitation spectra of heavy quark-antiquark bound states. In many of these instances, one has to deal with energy eigenstates of the isotropic harmonic oscillator in 3 dimensions that are simultaneously classified by their angular momentum. For example, the famous $J/\psi$ meson is a bound state of a charm quark $c$ and a charm antiquark $\bar c$ with spin $j=1$. The two valence quarks in the $J/\psi$ are in a spin $s=1$ configuration, and their orbital angular momentum is $l=0$ \cite{ParticleDataGroup:2020ssz}. It is useful for applications in nuclear and particle physics to compute the probability for a given $c-\bar c$ two-particle state to coalesce into a $J/\psi$ bound state, and to distinguish this process from the coalescence into other mesons, such as a $\chi_{c2}$, which is a $c\bar c$ bound state with $l=1$ and $j=1$. Such coalescence, or recombination, processes of quarks have found applications in the description of hadronization, which is the process for quarks and gluons to form bound states due to confinement of the strong force \cite{Fries:2008hs,Fries:2004ej,Fries:2003vb,Fries:2003kq,Greco:2003xt,Greco:2003mm,Greco:2003vf,Nonaka:2003ew,Oh:2009zj,ExHIC:2010gcb}.

In this work, we want to lay out a general formalism to compute recombination probabilities of two particles into bound states with well-defined orbital angular momentum quantum number $l$, in which the interaction is given by an isotropic harmonic oscillator 
potential in 3 dimensions. Although this work is motivated by applications to meson bound states formed by quark-antiquark pairs, the formalism can be applied to other systems in which an isotropic harmonic oscillator potential presents a suitable approximation, e.g. the recombination of nucleons into light nuclei \cite{Sato:1981ez,Baltz:1995tv,Scheibl:1998tk,Mattiello:1996gq,Kahana:1996bw,Chen:2003qj,Chen:2003ava,Zhu:2015voa,Zhao:2018lyf} and the recombination of hadrons into hadronic molecules \cite{Nonaka:2003ew,ExHIC:2011say,ExHIC:2017smd}.  The observables computed and approximations made in those cases in nuclear and particle physics differ somewhat from applications of recombination in atomic and plasma physics, where the focus is typically on the photon radiation emitted by these processes. Photons are often the primary observables for the latter, in particular for recombination into atoms in the early Universe where they form the Cosmic Microwave Background. The de-excitation of electrons in Coulomb-like potentials through the absorption and emission of photons plays a fundamental role and has been discussed in detail since the 1950s \cite{BetheSalpeter1957,1961ApJS....6..167K}. In contrast, in the formation of hadrons and nuclei in nuclear reactions, the recombined final states themselves are of interest. Moreover, electromagnetic final-state radiation is suppressed due to the relatively small value of the electromagnetic fine structure constant compared to the strong coupling constant. Energy and momentum conservation in $2\to 1$ and $3\to 1$ recombination processes has to be rather provided by spectators, although the detailed modeling of this aspect is often omitted.

For this work, we will assume that the information available about the initial two particles are the average positions and momenta at a fixed point in time. If no other information is known, the minimum assumption to make is that both particles are given by Gaussian wave packets around those average values with certain widths~\footnote{For the particular applications in nuclear physics we have in mind, this limited knowledge of the initial state usually applies. Expectation values for positions and momenta of particles are computed in classical approximations, e.g.\ Boltzmann-type transport, with little additional information available.}. The nature of the information given as input leads one naturally to consider a phase-space formalism for the calculation. Therefore, we revisit the Wigner distribution functions of angular momentum eigenstates of the isotropic 3-D harmonic oscillator. These were first calculated, to our knowledge, for a different application in nuclear physics by Prakash and Shlomo~\cite{Shlomo:1981ayz}.  A few special cases of Wigner functions with low orbital angular momentum have also been given in Refs.~\cite{Baltz:1995tv,ExHIC:2011say,PhysRevC.91.054914}. While Prakash and Shlomo used Moshinsky brackets in their derivation \cite{Moshinsky:1959qbh}, we will take a different path. Since the Wigner distributions for the 1-D harmonic oscillator are well known, see e.g.\ Refs.\ \cite{Groenewold1946,Curtright:2000ux}, we expand the 3-D angular momentum eigenstates in terms of states factorized into 1-D eigenstates. This allows us to compute the 3-D Wigner distributions from their 1-D counterparts. Generally, Wigner distributions of angular momentum eigenstates in 3-D seem to have not received much attention in the literature.  An exception, to some extent, are eigenstates in the Coulomb potential, see e.g.\ Refs.\ \cite{doi:10.1080/00268978200100752,Praxmeyer_2006}. However, elegant analytic expressions have been found
for angular momentum eigenstates of the harmonic oscillator in the 2-D case by Simon and Agarwal in Ref.\ \cite{Simon:00}.

The coalescence of two particles into energy eigenstates without well-defined orbital angular momentum was already discussed by some of us in \cite{Han:2016uhh}. We had found that for a particular ratio of initial wave packet widths to the width of the harmonic oscillator potential, the probability is simply a Poisson distribution in the energy quantum number $N$ of the bound state. We will confirm this result. Moreover, we will be able to see how the initial angular momentum $\mathbf{L}$ of the coalescing particles determines the partition of probabilities into excited states with the same energy $N$ but different orbital angular momentum quantum numbers $l$. Utilizing a phase-space formulation of quark coalescence allows us to interpret our mathematical results quite intuitively. The preference for particular final-state quantum numbers connects
directly to the relation between initial relative average momenta and positions of the coalescing particles. 
For the rest of this manuscript, angular momentum will refer to orbital angular momentum. We will assume the two particles to have spin 0 and to be distinguishable. However, our results will apply to initial particles of any spin as long as the spin-orbit coupling can be neglected~\footnote{For the coalescence of spin-$1/2$ quarks, this assumption is usually made.}.

Our work presents three main results. In the next section, we derive general expressions for the expansion coefficients of 3-D angular momentum eigenstates of the harmonic oscillator in terms of eigenstates factorized into 1-D eigenstates. In Sec.\ \ref{sec:3}, we use these coefficients to compute the Wigner distributions for angular momentum eigenstates of given quantum number $l$ and averaged over the magnetic quantum number $m$~\footnote{The averaging over $m$ can be easily left out but leads to considerably lengthier results. They are not needed for the applications we have in mind (the particular $j_3$ of a meson is not measured in most experiments) and we do not present them here. Ref.\ \cite{Shlomo:1981ayz} makes the same choice for their applications.}. We then compute the coalescence probabilities for two Gaussian wave packets in Sec.\ \ref{sec:4}. Finally, we discuss the implications of our results in Sec.\ \ref{sec:5}.


\section{Angular Momentum Eigenstates}
\label{sec:2}

\subsection{Conventions}

Let us begin by stating our conventions and notations. We consider the isotropic 3D-harmonic oscillator with Hamiltonian
\begin{equation}
  H=\frac{\mathbf{p}^2}{2 m} + \frac{1}{2} m \omega^2 \mathbf{r}^2  ,
\end{equation}
where $\mathbf{p}$ and $\mathbf{r}$ are the usual momentum and position operators, $m$ is the mass (the reduced mass in case of a 2-body system described by a harmonic oscillator potential), and $\omega$ describes the strength of the potential. It will be helpful to briefly introduce our conventions for the 1-D harmonic oscillator before we proceed. With the corresponding 1-D Hamiltonian $H=p^2/2 m + \omega^2 x^2/2$, we will use the set of orthonormal eigenfunctions
\begin{equation}
  \label{eq:1dho}
  \phi_n (x) = \sqrt{\frac{\nu}{2^n n! \sqrt{\pi}}}  H_n \left(\nu x\right) e^{-\frac{\nu^2 x^2}{2}}  \, ,
\end{equation}
where eigenstates are labeled by integers $n \ge 0$, and the corresponding energies are $E_n = \hbar \omega \left( n+\frac{1}{2} \right)$. It will turn out to be convenient to introduce the inverse natural length scale\footnote{It will be unnecessary to denote mass by $m$ from here on,
and we will rather use this letter to denote the magnetic quantum number.} of the oscillator $\nu =\sqrt{m\omega/\hbar}$. The $H_n$ are the usual Hermite polynomials.

Returning to the 3-D case, we note that two sets of energy eigenfunctions are usually used in the literature. 
\\ 
(a) \emph{Factorized eigenstates} (FE) utilize the factorizability of the Hamiltonian in the three cartesian coordinates, leading to eigenfunctions that are products of the 1-D
harmonic oscillator eigenfunctions
\begin{equation}
  \Phi_{n_1 n_2 n_3}(x,y,z) = \phi_{n_1}(x) \phi_{n_2}(y) \phi_{n_3}(z)  \, .
	\label{eq:fe}
\end{equation}
Here, the three integer quantum numbers $n_i \ge 0$ ($i=1,2,3$) label the eigenstates
with energies
\begin{equation}
  E_{n_1n_2n_3} = \hbar \omega \left(n_1+n_2+n_3 + \frac{3}{2} \right)   \, .
\end{equation}
(b) \emph{Angular momentum eigenstates} (AME) are simultaneous eigenstates to $H$, $\mathbf{L}^2$ and $L_3$, where $\mathbf{L}=(L_1,L_2,L_3)$ is the angular momentum operator. They can be expressed in spherical coordinates $(r,\theta,\phi)$ as
\begin{equation}
   \Psi_{klm}(r,\theta,\phi) = \sqrt{\frac{\nu^3 2^{k+l+2} k!}{\sqrt{\pi}(2k+2l+1)!!}} (\nu r)^l e^{-\frac{\nu^2 r^2}{2}} 
   L_k^{\left(l+\frac{1}{2}\right)}\left( \nu^2 r^2\right) Y_l^{m} (\theta,\phi) \, ,
\end{equation}
where the $L_k^{\left( l+\frac{1}{2}\right)}$ are associated (or generalized) Laguerre polynomials and the $Y_{l}^m(\theta,\phi)=Y_{l}^m(\hat{\mathbf r})$ are spherical harmonics, with $\hat{\mathbf{r}}=\mathbf{r}/r$ denoting the unit position vector. The energy of a state is given by the integer radial and angular momentum quantum numbers $k, l \ge 0$ as 
\begin{equation}
  E_{klm} = \hbar \omega \left( 2k+l + \frac{3}{2} \right)  \, .
\end{equation}
As usual, the magnetic quantum number $m$ is an integer and is bounded such that $-l \le m \le l$.

\subsection{Expansion of Angular Momentum Eigenstates in terms of Factorized Eigenstates}

It will be useful to express angular momentum eigenstates in terms of factorized
eigenstates to utilize the enormous amount of results available for the 1-D harmonic oscillator in the literature. Therefore, we want to find the expansion coefficients $C_{klm,n_1 n_2 n_3}$ in the equation
\begin{equation}
  \Psi_{klm}(\mathbf{r}) = \sum_{n_1 n_2 n_3} C_{klm,n_1 n_2 n_3} \Phi_{n_1 n_2 n_3}(\mathbf{r})  \, .
  \label{eq:expansion}
\end{equation}
As we are dealing with complete and properly normalized sets of states, we have the explicit expression
\begin{equation}
  C_{klm,n_1 n_2 n_3} = \int d^3  \mathbf{r}\, \Phi_{n_1 n_2 n_3}^*(\mathbf{r}) \Psi_{klm}(\mathbf{r})   \, .
  \label{eq:cintegral}
\end{equation}
It is straightforward to compute these integrals for any given small values for the sets of quantum numbers, see Table \ref{tab:c}.
We have not found a general expression for these coefficients in the literature, so we dedicate the remainder of this section to compute it.

Let us first note that the sum in Eq.\ (\ref{eq:expansion}) is restricted to the subspace of degenerate factorized eigenstates with the same energy eigenvalue $E=\hbar \omega (N+\frac{3}{2})$ as the left hand side. Here $N \equiv n_1+n_2+n_3 = 2k + l $ is the integer energy quantum number, and the dimension of the corresponding energy-degenerate
subspace is $d_N = \frac{1}{2} (N+1)(N+2)$.

The value of the general integral in Eq.\ (\ref{eq:cintegral}) can be readily obtained from a more general integral that involves replacing the special functions in the integrand with their generating functions. This will lead to integrals to be Gaussian and exponential in nature and the difficulty is shifted to taking the correct derivatives to recover the original expressions. We will use the well-known generating functions for Hermite and associated Laguerre polynomials,
\begin{align}
  h(t, u) &= 
   e^{-t^2 + 2 u t} = \sum_{n=0}^\infty H_n(u) \frac{t^n}{n!}   \, , \\
  l^{(\alpha)} (s,u) &= 
  {\left( 1-s \right) }^{-\alpha-1} e^{-\frac{s u}{1-s}} = 
  \sum_{n=0}^\infty L^{(\alpha)}_n(u) s^n   \, ,
\end{align}
as well as a variant of the Herglotz generating function for spherical harmonics \cite{CourantHilbert1953},
\begin{equation}
  y(v,\lambda ; {\mathbf r}) = 
   e^{v \mathbf{a} \cdot  {\mathbf r}} 
   = \sum_{l=0}^\infty \sum_{m=-l}^l \sqrt{\frac{4\pi}{2l+1}}
  \frac{r^l v^l \lambda^m}{\sqrt{(l+m)!(l-m)!}} Y^m_l \left( \hat{\mathbf r} \right)  \, .
\end{equation}
In the above, we have defined the auxiliary vector
\begin{equation}
  \mathbf{a} = \left( -\frac{\lambda}{2} + \frac{1}{2\lambda} , -i \frac{\lambda}{2} - i \frac{1}{2\lambda}, 1 \right)  \, ,
\end{equation}
which satisfies $\mathbf{a}^2=0$.  The spherical harmonics can be recovered from this Laurent series by means of
\begin{equation}
   Y^m_l \left(\hat{\mathbf{r}} \right) = \sqrt{\frac{2l+1}{4\pi}} \sqrt{\frac{(l-m)!}{(l+m)!}}  
   \frac{1}{l!} \frac{\partial^l}{\partial v^l} \frac{\partial^{l+m}}{\partial \lambda^{l+m}} \left[
   \left( \frac{\lambda}{r}\right)^l  y(v,\lambda; {\mathbf r}) 
   \right]_{\substack{\lambda=0 \\ v=0}}
    \, ,
\end{equation}
similar to the formulas to recover the Hermite and associated Laguerre polynomials from their respective series,
\begin{equation}
  H_n(u)=\frac{\partial^n h(t,u)}{\partial t^n}\Bigg|_{t=0} \, ,  
  \qquad
  L_n^{(\alpha)}(u)=\frac{1}{n!}\frac{\partial^n l^{(\alpha)}(s,u)}{\partial s^n}\Bigg|_{s=0} \, . 
\end{equation}

By replacing each special function implicit in Eq.\ (\ref{eq:cintegral}) by its generating function, $H_{n_i} \to h$, $L_k^{(\alpha)} \to l^{(\alpha)}$, $Y_l^m \to r^{-l} y$, we define a generating function $\mathcal{F}$ for the coefficients in Eq.\ (\ref{eq:cintegral}) as
\begin{multline}
	\label{eq:fdef}
  \mathcal{F}(t_1,t_2,t_3,s,v,\lambda)  = \sqrt{\frac{2^{k+l+2-n_1-n_2-n_3}k!}{n_1!n_2!n_3!(2k+2l+1)!! }} 
  \frac{\nu^{l+3}}{\pi}  (1-s)^{-l-\frac{3}{2}}    \\ \times \int d^3 \mathbf{r} \,  e^{-\nu^2 r^2}  
  e^{-(t^2-2\nu\mathbf{t}\cdot \mathbf{r})} e^{-\frac{s\nu^2r^2}{1-s}} e^{v\mathbf{a}\cdot\mathbf{r}}
  \, .
\end{multline}
We have introduced the vector $\mathbf{t} = (t_1,t_2,t_3)$ of auxiliary variables used for the generating functions of the Hermite polynomials. The coefficients $C_{klm,n_1 n_2 n_3}$ can then be generated from $\mathcal{F}$ as derivatives with respect to the auxiliary variables
\begin{multline}
  \label{eq:cderiv}
  C_{klm,n_1 n_2 n_3} = \sqrt{\frac{2l+1}{4\pi}} \sqrt{\frac{(l-m)!}{(l+m)!}}\frac{1}{k!l!}   \\
  \times
   \frac{\partial^{n_1+n_2+n_3+k+2l+m}}{\partial t_1^{n_1} \partial t_2^{n_2} t_3^{n_3} \partial s^k \partial v^l
   \partial \lambda^{l+m}} \lambda^l\cal F \Bigg|_{\substack{t_1 = t_2=t_3=0\\s=v=\lambda=0}}
    \, .
\end{multline}

To compute the integral in $\cal F$, we define the displacement vector
\begin{equation}
  \mathbf{U} = \frac{1-s}{\nu^2} \left(\nu\mathbf{t} + \frac{v}{2}\mathbf{a}  \right)  \, ,
\end{equation}
and rewrite the integrand in the last line of Eq.\ (\ref{eq:fdef}) as
\begin{equation}
   e^{-\frac{\nu^2}{1-s}\left( \mathbf{r}-\mathbf{U}\right)^2}  
  e^{-t^2+\frac{\nu^2}{1-s}U^2} 
  \, .
\end{equation}
 This is a Gaussian in $\mathbf{r}$, and the integral in Eq.\ (\ref{eq:fdef}) can now be readily evaluated to give a closed expression for the generating function of the expansion coefficients
\begin{equation}
  \label{eq:cgen}
   \mathcal{F}(t_1,t_2,t_3,s,v,\lambda) 
   =  \sqrt{ \frac{\pi 2^{k+l+2-n_1-n_2-n_3}k!}{n_1!n_2!n_3!(2k+2l+1)!! }} 
  \frac{\nu^{l}}{(1-s)^l}  e^{-t^2+\frac{\nu^2}{1-s}U^2} 
  \, .
\end{equation}

Next, we have to execute the derivatives in Eq.\ (\ref{eq:cderiv}). We note that, since $\mathbf{a}^2=0$, the square of the displacement vector 
$\mathbf{U}$ is a linear function in the auxiliary variable $v$,
\begin{equation}
  \mathbf{U}^2 = \frac{(1-s)^2}{\nu^2} t^2 + \frac{(1-s)^2}{\nu^3} v \, \mathbf{t}\cdot \mathbf{a}  \, .
\end{equation}
The $l$ derivatives with respect to $v$ thus generate a factor $(1-s)^l(\mathbf{t}\cdot \mathbf{a})^l/\nu^l $ in Eq.\ (\ref{eq:cgen}), which cancels all powers of $(1-s)$ outside of the exponential. Subsequently setting $v=0$ simplifies the derivatives with respect to $s$ to derivatives of 
$\exp[-st^2]$, which simply generates $k$ powers of $-t^2$. Hence, we have
\begin{multline}
  \label{eq:c1}
  C_{klm,n_1 n_2 n_3}  = \frac{(-1)^k}{l!} \sqrt{\frac{(l-m)!}{(l+m)!}}  
  \sqrt{ \frac{(2l+1) 2^{k-l-n_1-n_2-n_3}}{n_1!n_2!n_3! k!(2k+2l+1)!! }}   \\ \times
  \frac{\partial^{n_1+n_2+n_3+l+m}}{\partial t_1^{n_1} \partial t_2^{n_2} t_3^{n_3}
  \partial \lambda^{l+m}} Q^l t^{2k} \Bigg|_{\substack{t_1 = t_2=t_3=0\\ \lambda=0}}  
   \, ,
\end{multline}
where we have defined 
\begin{equation}
   Q = 2 \lambda\, \mathbf{t} \cdot \mathbf{a} =
   t_1(1-\lambda^2) - it_2 (1+\lambda^2) + 2t_3 \lambda\, .
\end{equation}

Although not completely explicit, Eq.\ (\ref{eq:c1}) delivers a rather compact notation for the expansion coefficients. It is well suited for quick analytic computations of 
coefficients for arbitary $k$, $l$. 
Table \ref{tab:c} gives the complete set of coefficients up to  $N=2k+l=2$.

\begin{table}[t]
  \begin{center}
  \def\arraystretch{1.0}
  {\begingroup\makeatletter\def\f@size{9}\check@mathfonts
	\begin{tabular}{|c||c||c|c|c|c|c|c|}
  \hhline{|-||--|~~~~~}
  \multirow{3}{*}{$N=0$} &
     \multicolumn{2}{|l|}{$k=0$  \quad $l =0$} & \multicolumn{5}{c}{} \\
     \hhline{|~||==|~~~~~}
     & $(n_1,n_2,n_3)$ &  $(0,0,0)$ & \multicolumn{4}{c}{}\\ \cline{2-3}
     &  $m=0$  & $1$  &\multicolumn{4}{c}{} \\
   \hhline{|=||==|--~~~}
  \multirow{3}{*}{$N=1$} &
     \multicolumn{4}{|l|}{$k=0$  \quad $l =1$} & \multicolumn{3}{c}{}  \\
     \hhline{|~||==|==|~~~}
     &  $(n_1,n_2,n_3)$ & $(1,0,0)$ & $(0,1,0)$ & $(0,0,1)$ & \multicolumn{3}{c}{}  \\  \cline{2-5}  
     &   {$m=1$}  & $-\frac{1}{\sqrt{2}}$ & $\frac{i}{\sqrt{2}}$ & $0$  & \multicolumn{3}{c}{} \\  \cline{2-5}
     &  {$m=0$}   & $0$ & $0$ & $1$ & \multicolumn{2}{c}{}  \\ \cline{2-5}
		&  {$m=-1$}   & $\frac{1}{\sqrt{2}}$ & $\frac{i}{\sqrt{2}}$ & $0$  \\ 
   \hhline{|=||==|==---}
  \multirow{12}{*}{$N=2$} &
     \multicolumn{7}{|l|}{$k=0$  \quad $l =2$}  \\
     \hhline{|~||==|=====|}
     &  $(n_1,n_2,n_3)$  & $(2,0,0)$ & $(1,1,0)$ & $(0,2,0)$ & $(0,1,1)$ & $(0,0,2)$ & $(1,0,1)$ \\    \cline{2-8}  
     & {$m=2$}  & $\frac{1}{2}$ & $ -\frac{i}{\sqrt{2}}$ & $-\frac{1}{2}$ & $0$ &  $0$ & $0$  \\  \cline{2-8}
     & {$m=1$}  & $0$ & $0$ & $0$ & $\frac{i}{\sqrt{2}}$ & $0$ & $-\frac{1}{\sqrt{2}}$ \\ \cline{2-8}
		&  {$m=0$}  & $-\frac{1}{\sqrt{6}}$ & $0$ & $-\frac{1}{\sqrt{6}}$ & $0$ & $\sqrt{\frac{2}{3}}$  & $0$ \\  \cline{2-8}
 		&  {$m=-1$}  & $0$ & $0$ & $0$ & $\frac{i}{\sqrt{2}}$ & $0$ & $\frac{1}{\sqrt{2}}$ \\ \cline{2-8}
		&  {$m=-2$}  & $\frac{1}{2}$ & $ \frac{i}{\sqrt{2}}$ & $-\frac{1}{2}$ & $0$ &  $0$ & $0$ \\ 
    \hhline{|~||==|=====|}
    &  \multicolumn{7}{|l|}{$k=1$  \quad $l =0$}  \\
     \hhline{|~||==|=====|}
     &  $(n_1,n_2,n_3)$  & $(2,0,0)$ & $(1,1,0)$ & $(0,2,0)$ & $(0,1,1)$ & $(0,0,2)$ & $(1,0,1)$ \\    \cline{2-8}  
     &  {$m=0$}  & $-\frac{1}{\sqrt{3}}$ & $0$ & $-\frac{1}{\sqrt{3}}$ & $0$ & $-\sqrt{\frac{1}{3}}$  & $0$ \\  
  \hline
  \end{tabular}
  \endgroup}
  \caption{\label{tab:c} Coefficients $C_{klm,n_1 n_2 n_3}$ appearing in the expansion given in Eq.\ (\ref{eq:expansion}), for the lowest energy eigenstates.}
  \end{center}
\end{table}

\subsection{Further Discussion of the Expansion Coefficients}

Although carrying out the derivatives in Eq.\ (\ref{eq:c1}) for the general case leads to tedious expressions, it reveals useful constraints that severely restrict the number of coefficients that are non-zero. We quote the two most important constraints at the beginning of this subsection and will then proceed to justify them. We find that the $C_{klm,n_1 n_2 n_3}$ are non-zero only if 
\begin{align}
\label{eq:constr1}
2k+l  &= n_1+n_2+n_3   \\
& \text{and}   \notag \\
\label{eq:constr2}
l+m-n_3 &= 0\, \text{mod}\, 2 
    \, .
\end{align}
The first constraint simply reflects the fact that the energy eigenvalues of the functions involved on the left-hand and right-hand sides in Eq.\ (\ref{eq:expansion}) need to be equal. The second constraint, forcing $l+m-n_3$ to be even, is less obvious but useful. It sets about half of the remaining coefficients to be zero.

To compute the derivatives with respect to components of $\mathbf{t}$, we apply the Binomial Theorem and Leibniz's product rule several time to expand the derivatives of the term $Q^l t^{2k}$. We can resolve several of these sums after setting $\mathbf{t}=0$. The remaining sums read
\begin{multline}
    \frac{\partial^{n_1+n_2+n_3}}{\partial t_1^{n_1} \partial t_2^{n_2} t_3^{n_3}} Q^l t^{2k} \Bigg|_{\mathbf{t}=\mathbf{0}}  
   = k! l!  \delta_{n_1+n_2+n_3,2k+l} \sum_{\substack{j_1+j_2+j_3 = k \\ 2j_\mu \le n_\mu, \mu=1,2,3}} \begin{pmatrix}n_1\\2j_1\end{pmatrix}
    \begin{pmatrix}n_2\\2j_2\end{pmatrix}\begin{pmatrix}n_3\\2j_3\end{pmatrix} \\ \times
     \frac{(2j_1)!(2j_2)!(2j_3)!}{j_1!j_2!j_3!} {(1-\lambda^2)}^{n_1-2j_1} {(-i (1+\lambda^2))}^{n_2-2j_2} {(2\lambda)}^{n_2-2j_3}
   \, .
\end{multline}
This is where energy conservation emerges as a constraint, ensured by the Kronecker-$\delta$ in the expression above. 
After taking the remaining derivatives with respect to $\lambda$, the expansion coefficients are
\begin{multline}
  \label{eq:finalcoeff1}
  C_{klm,n_1 n_2 n_3} = (-1)^k \sqrt{ (2l+1) 2^{k-l-n_1-n_2-n_3} } \\ \times \sqrt{\frac{n_1!n_2!n_3! k! (l-m)! (l+m)!}{(2k+2l+1)!! }} \mathcal{S}\, .
\end{multline}  
The expression $\mathcal S$ contains the residual sums
\begin{multline}
\mathcal{S} = \sum_{\substack{j_1,j_2,j_3=0 \\ j_1+j_2+j_3 = k}}^{n_1, n_2,n_3} \frac{2^{n_3-2j_3} i^{n_2-2j_2}}{j_1!j_2!j_3! (n_1-2j_1)! (n_2-2j_2)! (n_3-2j_3)!}
    \\ \times \sum_{\rho=0}^{\kappa+j_3} (-1)^{\rho}\begin{pmatrix}n_1-2j_1 \\ \rho\end{pmatrix}\begin{pmatrix}n_2-2j_2 \\ \kappa+j_3-\rho\end{pmatrix}\, ,
\end{multline}
with additional constraints on the indices $j_1$ and $j_2$ in the sum as $2j_1 \ge n_1-l$  and $2j_1+2j_2 \ge n_1+n_2-l$, and overall constraints as given in Eqs.\ (\ref{eq:constr1}) and (\ref{eq:constr2}). In the above, we have used the shorthand notation
\begin{equation}
  \kappa = \frac{1}{2} \left( l+m-n_3 \right) \, ,
\end{equation}
which is an integer due to Eq.(\ref{eq:constr2}). The sum over $\rho$ in $\mathcal{S}$
represents a value of the Gaussian hypergeometric function ${}_2F_1$ at the point $-1$. To be specific, we have
\begin{multline}
   \label{eq:finalcoeff2}
\mathcal{S} = \sum_{\substack{j_1,j_2,j_3=0 \\ j_1+j_2+j_3 = k}}^{n_1, n_2,n_3} \frac{2^{n_3-2j_3} i^{n_2-2j_2}}{j_1!j_2!j_3! (n_1-2j_1)! (n_2-2j_2)! (n_3-2j_3)!}
    \\ \times \begin{pmatrix}n_2-2j_2 \\ \kappa+j_3 \end{pmatrix} {}_2F_1(-\kappa-j_3,-n_1+2j_1;1-\kappa-j_3+n_2-2j_2;-1)  \, .
\end{multline}

Eqs.\ (\ref{eq:finalcoeff1}) and (\ref{eq:finalcoeff2}) with the additional constraints given constitute the main result of this subsection. A closed expression can be given 
in the important case $k=0$ as
\begin{multline}
    \label{eq:finalcoeff3}
    C_{0lm,n_1 n_2 n_3} =  \sqrt{\frac{ (l+m)! (l-m)! }{2^{2l} n_1! n_2!n_3! (2k+2l-1)!! }} \\ \times 
    2^{n_3}i^{n_2} \begin{pmatrix}n_2 \\ \kappa \end{pmatrix} {}_2F_1(-\kappa,-n_1;1-\kappa+n_2;-1) \, .
\end{multline}

\section{Wigner Representation of Angular Momentum Eigenstates}
\label{sec:3}

\subsection{Wigner Distributions and Review of the 1-D Case}

We define the generalized Wigner transformation, applied to a wave function $\psi_1(\mathbf{r})$ and a complex conjugate wave function $\psi_2^*(\mathbf{r})$. The transformation yields the phase-space distribution 
\begin{equation}
  W_{\psi_2, \psi_1} (\mathbf{r},\mathbf{q}) = \int \frac{d^3 \mathbf{r}'}{(2\pi \hbar)^3} e^{\frac{i}{\hbar}\mathbf{r'}\cdot\mathbf{q}}  \,
  \psi_2^*\left( \mathbf{r} + \frac{1}{2}\mathbf{r'} \right) 
  \psi_1 \left(  \mathbf{r} - \frac{1}{2}\mathbf{r'} \right) \, ,
  \label{eq:wignerdef}
\end{equation}
which in general can take complex values. We recover ordinary, real-valued Wigner distributions for the special case 
$\psi_1=\psi_2$. 

We use corresponding definitions in the 1-D case. Using the short-hand notation $W_{n' \, n} = W_{\phi_{n'}, \phi_{n}}$, the Wigner distributions for the energy eigenstates of the 1-D harmonic oscillator from Eq.\ (\ref{eq:1dho}) are \cite{Groenewold1946,Curtright:2000ux}
\begin{equation}
   W_{n' \, n}(x,q) =  \frac{(-1)^{n'}}{\pi \hbar} \sqrt{\frac{n'}{n}} 
   u^{\frac{n-n'}{2}} e^{-u/2} e^{-i(n-n') \zeta} L_{n'}^{(n-n')}(u) \, ,
\end{equation}
where $u= 2 \left( q^2/(\hbar^2\nu^2) + \nu^2 x^2 \right)$ and $\tan \zeta = q/(\hbar \nu^2 x)$.
Deviating from Ref.\ \cite{Curtright:2000ux}, we are keeping physical units. The formulas have been adjusted accordingly. The diagonal Wigner functions $W_{nn}$ for the 1-D harmonic oscillator are included as a special case \cite{Han:2016uhh}. 
The lesser known $W_{n' \, n} $ for $n\ne n'$ are the off-diagonal Wigner distributions for the harmonic oscillator energy eigenstates.

Later, we will utilize the generating function for these generalized Wigner functions, which is discussed at length in Ref.\ \cite{Curtright:2000ux},
\begin{equation}
  \label{eq:wgen}
  G(\alpha,\beta; x,q) = \frac{1}{\pi\hbar} e^{\alpha\beta - \left( \nu x-\frac{\alpha+\beta}{\sqrt{2}}\right)^2 - \left( \frac{q}{\hbar \nu}
  + i\frac{\alpha-\beta}{\sqrt{2}}\right)^2}  \, .
\end{equation}
We recover the Wigner functions as
\begin{equation}
   \label{eq:wgen2}
   W_{n' \, n} (x,q)= \frac{1}{\sqrt{n! n'!}}\frac{\partial^{n+n'}}{{\partial \alpha^{n'}}{\beta^{n}}}G(\alpha,\beta;x,q)\Bigg|_{\alpha=\beta=0}
   \, .
\end{equation}
With these results from the literature lined up, we are now ready to compute the Wigner distributions for the isotropic 3-D harmonic oscillator using the expansion coefficients from the previous section.

\subsection{Wigner Distributions in 3-D}
\label{sec:3.2}

We define the shorthand notation $W_{klm} = W_{\Psi_{klm}, \Psi_{klm}}$ for the diagonal Wigner distributions of energy and angular momentum eigenstates. If one is not interested in the magnetic quantum number, e.g.\ if the polarization of a bound state in the harmonic oscillator is not considered, one would rather like to consider the $m$-averaged distributions
\begin{equation}
  W_{kl} = \frac{1}{2l+1}\sum_{-l \le m \le l} W_{klm}  \, .
\end{equation}
The $m$-dependent distributions are easy to compute but difficult to visualize due to a lack of symmetries. We will focus on the $m$-averaged distributions in the following as the applications we have in mind usually do not require knowledge of the angular
momentum projection. Combining Eqs.\ (\ref{eq:wignerdef}), (\ref{eq:expansion}), and (\ref{eq:fe}) with the results of the previous subsection, we see that
\begin{equation}
   W_{kl}(\mathbf{r}, \mathbf{q}) =  \! \sum_{\substack{n_1,n_2,n_3 \\ n'_1,n'_2,n'_3}}  D_{kl}\! \left( \substack{n_1,n_2,n_3 \\ n'_1,n'_2,n'_3} \right)
  W_{n'_1 n_1}(r_1,q_1) W_{n'_2 n_2}(r_2,q_2) W_{n'_3 n_3}(r_3,q_3)   \, .
   \label{eq:wignerferep}
\end{equation}
The expansion coefficients for Wigner distributions of $m$-averaged angular momentum eigenstates in terms of their factorized counterparts are
\begin{equation}
    D_{kl}\! \left( \substack{n_1,n_2,n_3 \\ n'_1,n'_2,n'_3} \right)= \frac{1}{2l+1}\sum_m c^*_{klm,n'_1 n'_2 n'_3}c_{klm,n_1 n_2 n_3}  \, .
\end{equation}
The constraints (\ref{eq:constr1}) and (\ref{eq:constr2}) immediately put constraints on these coefficients, which we do not spell out here in detail.

\begin{figure}[t]
\begin{center}
  \includegraphics[width=0.3\columnwidth]{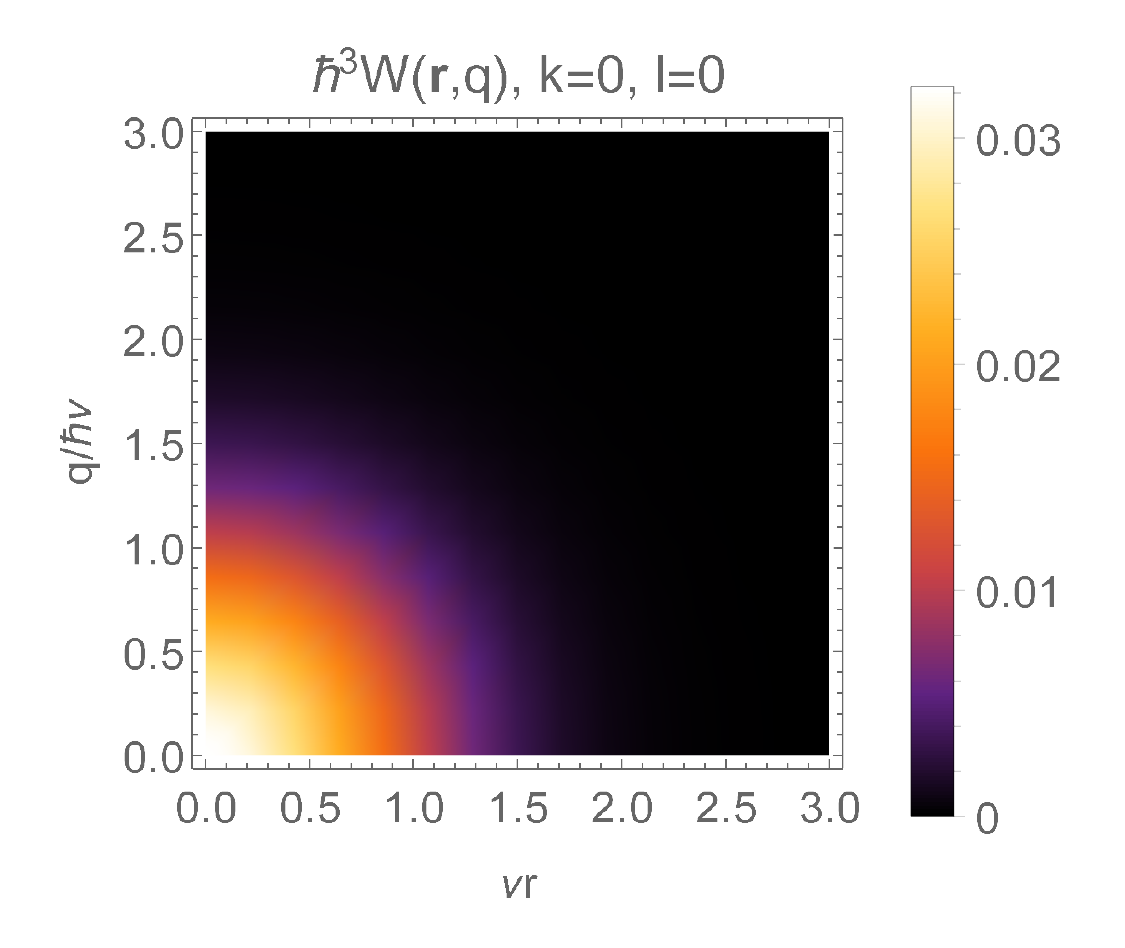}~ 
  \includegraphics[width=0.3\columnwidth]{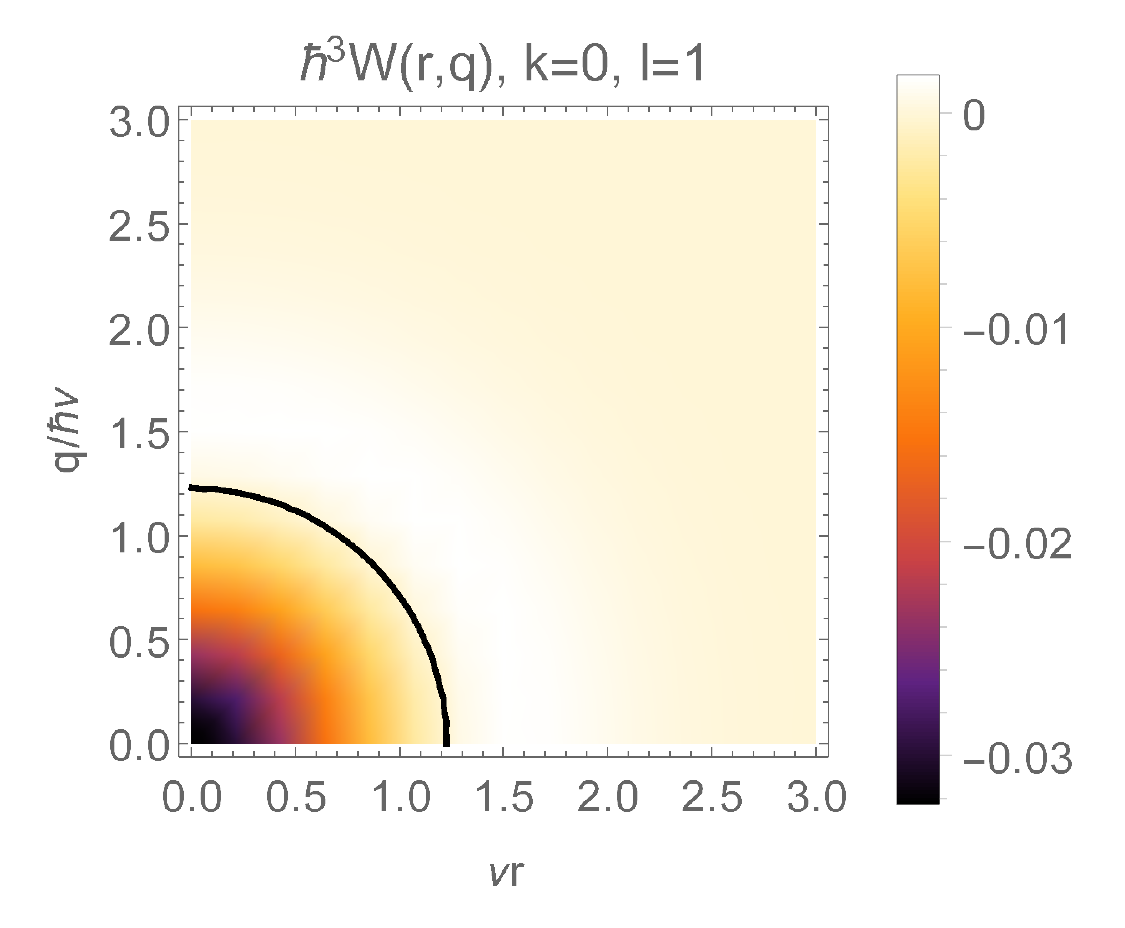} \\
  \includegraphics[width=1.0\textwidth]{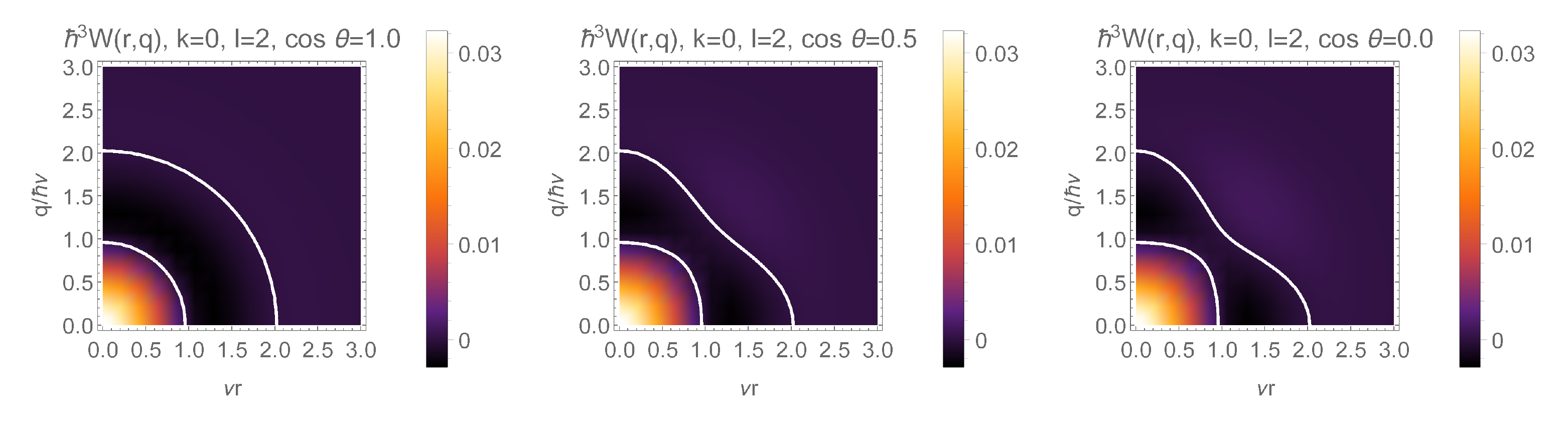} \\
  \includegraphics[width=1.0\textwidth]{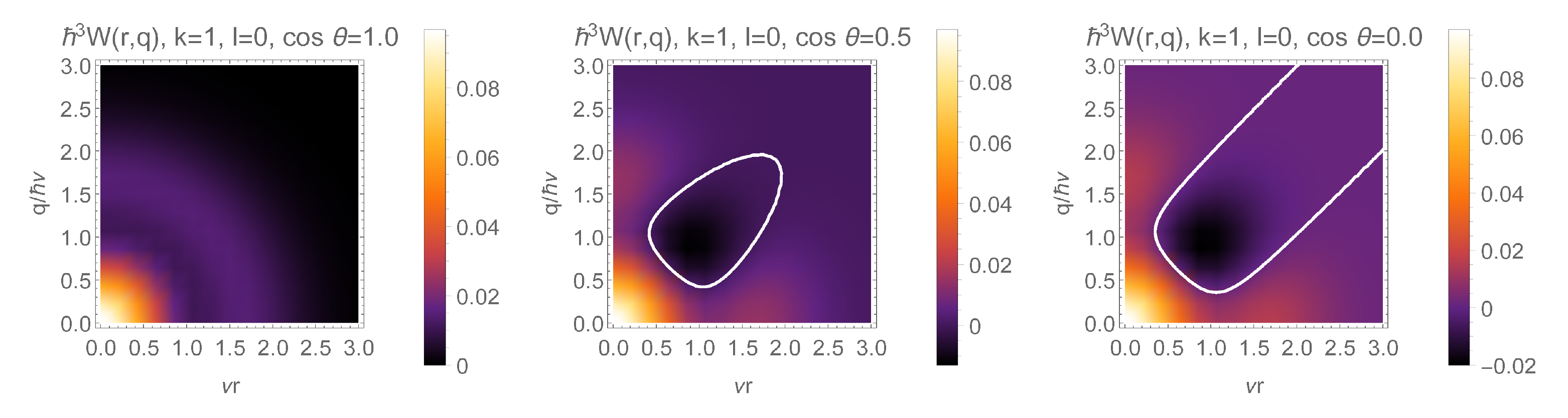}
\caption{\label{fig:wplots} Phase-space distributions $W_{kl}$ for $m$-averaged angular momentum eigenstates for the lowest $k$- and $l$-states as functions of $r=|\mathbf{r}|$ and $q=|\mathbf{q}|$. The thick black or white lines indicate nodes where $W_{kl}=0$. Distributions that depend on the scalar product $\mathbf{r}\cdot \mathbf{q}$ are plotted for several values of the angle $\theta$ given by $\cos\theta = \mathbf{r}\cdot \mathbf{q}/rq$.}
\end{center}
\end{figure}

Working out the algebra readily delivers the $m$-averaged distributions. We tabulate them here up to $N=3$:
\begin{align}
  \label{eq:wigfinal}
  W_{00} & =  \frac{1}{\pi^3\hbar^3} e^{-\frac{q^2}{\hbar^2 \nu^2}-\nu^2 r^2 }\, ,  \\   
  W_{01} & =  W_{00}
     \left( -1 +\frac{2}{3} \nu^2 r^2 +\frac{2}{3} \frac{q^2}{\hbar^2 \nu^2} \right)\, ,  \\
  W_{02} & =  W_{00}
     \left( 1 +  \frac{4}{15} \nu^4 r^4 -\frac{4}{3}\nu^2 r^2 +   \frac{16}{15} \frac{r^2 q^2}{\hbar^2} \right.  \\
    & \qquad   \left. 
     - \frac{8}{15} \frac{(\mathbf{r} \cdot \mathbf{q})^2}{\hbar^2}    
     - \frac{4}{3} \frac{q^2}{\hbar^2 \nu^2} +  \frac{4}{15} \frac{q^4}{\hbar^4 \nu^4} \right)\, ,   \\
  W_{10} & = W_{00}
     \left( 1 +  \frac{2}{3} \nu^4 r^4 -\frac{4}{3}\nu^2 r^2 \right.  \\
    & \qquad -   \left. \frac{4}{3} \frac{r^2 q^2}{\hbar^2} 
     + \frac{8}{3} \frac{(\mathbf{r} \cdot \mathbf{q})^2}{\hbar^2}        - \frac{4}{3} \frac{q^2}{\hbar^2 \nu^2} +  \frac{2}{3} \frac{q^4}{\hbar^4 \nu^4} \right)\, ,   \\
  W_{03} &= W_{00}
     \left( -1 +  \frac{8}{105} \nu^6 r^6 -  \frac{4}{5} \nu^4 r^4 + 2 \nu^2 r^2 \right.  \\
    & \qquad - \frac{16}{5} \frac{r^2 q^2}{\hbar^2}  \left( 1 - \frac{3}{14} \nu^2 r^2  -  \frac{3}{14}\frac{q^2}{\hbar^2 \nu^2} \right) \\
     & \qquad  + \frac{8}{5} \frac{(\mathbf{r} \cdot \mathbf{q})^2}{\hbar^2}     \left( 1 -  \frac{2}{7} \nu^2 r^2  
     -  \frac{2}{7}\frac{q^2}{\hbar^2 \nu^2} \right)   \\
    & \qquad\qquad    \left.+2 \frac{q^2}{\hbar^2 \nu^2} -  \frac{4}{5} \frac{q^2}{\hbar^4 \nu^4} +  
       \frac{8}{105} \frac{q^6}{\hbar^6 \nu^6}\right)\, ,   \\
  W_{11} &= W_{00}
     \left( -1 +  \frac{4}{15} \nu^6 r^6 -  \frac{22}{15} \nu^4 r^4 + 2 \nu^2 r^2 \right.  \\
    & \qquad +   \frac{4}{5} \frac{r^2 q^2}{\hbar^2}   \left( 1 -  \frac{1}{3} \nu^2 r^2  -  \frac{1}{3}\frac{q^2}{\hbar^2 \nu^2} \right) \\
     & \qquad  - \frac{56}{15} \frac{(\mathbf{r} \cdot \mathbf{q})^2}{\hbar^2}     \left( 1 -  \frac{2}{7} \nu^2 r^2  
     -  \frac{2}{7}\frac{q^2}{\hbar^2 \nu^2} \right)   \\
    & \qquad\qquad    \left.+2 \frac{q^2}{\hbar^2 \nu^2} -  \frac{4}{15} \frac{q^4}{\hbar^4 \nu^4} +  
       \frac{22}{15} \frac{q^6}{\hbar^6 \nu^6}\right)    \, .    \label{eq:wigfinal2}
\end{align}
The 3-D isotropic oscillator exhibits SO(3) symmetry together with a symmetry $\nu \mathbf r \leftrightarrow \mathbf{q}/(\hbar \nu)$. Both of these symmetries are also obeyed by the Wigner functions when averaged over $m$. The distributions $W_{kl}$  only depend on scalar products $r^2=|\mathbf{r}|^2$, $q^2=|\mathbf{q}|^2$ and $\mathbf{r}\cdot \mathbf{q}$. This makes their visualizations as functions of $r$, $q$ and $\theta$, the angle between the vectors $\mathbf{r}$ and $\mathbf{q}$, straightforward. We plot several $W_{kl}$ as functions of $r$ and $q$ in Fig.\ \ref{fig:wplots} for various values of $\theta$.
The Wigner distributions are well-behaved with no oscillations at large arguments $r$ and $q$, unlike those seen in the Coulomb-case even in the ground state \cite{doi:10.1080/00268978200100752,Praxmeyer_2006}.  Rather, the ground state distribution is positive definite, and the number of node lines increases slowly with $k$ and $l$.

Our results are consistent with those given for the cases of low values of $k$ and $l$ in Refs.\ \cite{Baltz:1995tv,ExHIC:2011say,PhysRevC.91.054914}. Also, Shlomo and Prakash \cite{Shlomo:1981ayz} have given expressions for the Wigner functions in the general case using Moshinsky brackets \cite{Moshinsky:1959qbh}. However, just as in our case, a closed expression is difficult to achieve. We have confirmed for a few explicit cases that our results coincide with their results as well.

\section{Coalescence Into Angular Momentum Eigenstates}
\label{sec:4}

\subsection{Differential and Total Coalescence Probabilities}

In the following, we want to describe the process of two quasi-free particles coalescing into a bound state described by the wave function of a 3-D isotropic harmonic oscillator potential. I.e., at short distances the particles are subject to forces approximated by those of a harmonic oscillator. Quasi-free here means that the particles initially will be assumed to be described by Gaussian wave packets centered around points 
$(\mathbf{r}_1,\mathbf{p}_1)$ and $(\mathbf{r}_2,\mathbf{p}_2)$ in phase space. To be precise, the Wigner distributions are assumed to have the form
\begin{equation} 
  W_{i} \left(\mathbf{x},\mathbf{k}\right) = \frac{1}{\pi^3\hbar^3 } e^{-\frac{(\mathbf{x}-\mathbf{r_i})^2}{2\delta^2} - 
   2\frac{\delta^2}{\hbar^2} (\mathbf{k}-\mathbf{p_i})^2 }
  \label{eq:gaussinput}
\end{equation}
for $i=1,2$. We assume both wave packets to be isotropic with common spatial width $\delta$. These latter restrictions can be lifted, in principle, if more information is known about the initial state particles. For the applications we have in mind, the full quantum information is usually poorly known and quasi-classical information, identified with the peaks of the Gaussians, together with a minimal smearing consistent with the uncertainty principle must often suffice.  

For the motion and position of the bound state after coalescence, we will assume a complete set of states representing plane waves with momentum $\mathbf{P}_f$. 
Applying the Wigner transformation Eq.\ (\ref{eq:wignerdef}) to a plane wave state of momentum $\mathbf{P}_f$, 
\begin{equation} 
  \psi_{\mathbf{P}_f} (\mathbf{r} ) = \frac{1}{( 2\pi \hbar)^{3/2}} e^{\frac{i}{\hbar} \mathbf{P}_f \cdot \mathbf{r}}\, ,
\end{equation}
yields the distribution
\begin{equation}
  \tilde W_{\mathbf{P}_f} (\mathbf{P}) = \frac{1}{(2\pi\hbar)^3} \delta^{(3)} \left( \mathbf{P} - \mathbf{P}_f \right)   \, .
\end{equation}
These distributions $\tilde W$ differ from ordinary Wigner distributions as they are not normalized to one, mirroring their wave function in that respect~\footnote{Unlike ordinary Wigner distributions, they are also not bounded.}.

In the Wigner formalism, the probability $\tilde \pcal_{klm,\mathbf{P}_f}d^3 \mathbf{P}_f$ for the coalescence process to result in a bound state
with momentum $\mathbf{P}_f$ and quantum numbers $k$, $l$, $m$ is given by the phase-space overlap integral between initial and final states,
\begin{multline}
   \tilde \pcal_{klm,\mathbf{P}_f} = 
   {(2\pi\hbar)}^6 \int d^3 \mathbf{x}_1 d^3 \mathbf{x}_2 d^3 \mathbf{k}_1 d^3 \mathbf{k}_2  \tilde W_{\mathbf{P}_f} (\mathbf{K})    \\ \times
   W_{klm}  \left( \Delta \mathbf{x} , \Delta \mathbf{k} \right) 
   W_1(\mathbf{x}_1,\mathbf{k}_1) W_2(\mathbf{x}_2,\mathbf{k}_2)   \, .
  \label{eq:coalp1}
\end{multline}
Here, $W_{klm}$ is the Wigner function of the harmonic oscillator bound state with quantum numbers $k,l,m$, and we have introduced the center and relative coordinates
\begin{align}
  \mathbf{X} &= \frac{1}{2}\left( \mathbf{x}_1 + \mathbf{x}_1\right) \, , & \Delta \mathbf{x} &= \mathbf{x}_1 - \mathbf{x}_2 
  \, , \\
  \mathbf{K} &= \mathbf{k}_1+\mathbf{k}_2  \, , &  \Delta \mathbf{k} \, , &= \frac{1}{2}\left( \mathbf{k}_1 - \mathbf{k}_2 \right)
  \, .
\end{align}

After recasting the integrals in Eq.\ (\ref{eq:coalp1}) into integrals over these center and relative coordinates, one can easily take the integrals over $\mathbf{X}$ and $\mathbf{K}$ to arrive at
\begin{multline}
\label{eq:pf0}
   \pcal_{klm,\mathbf{P}_f}=  8 J_{\mathbf{P}_i,\mathbf{P}_f}  e^{-\frac{r^2}{4\delta^2} - 4\frac{\delta^2}{\hbar^2} p^2}  \\ \times
    \int d^3 \Delta \mathbf{x} \, d^3 \Delta \mathbf{k}  \,
   W_{klm}\left( \Delta \mathbf{x} , \Delta \mathbf{k} \right)  
   e^{-\frac{\Delta x^2}{4\delta^2} + \frac{1}{2\delta^2} \Delta \mathbf{x}\cdot \mathbf{r}}
   e^{-4\frac{\delta^2}{\hbar^2} \Delta k^2 + 8 \frac{\delta^2}{\hbar^2} \Delta \mathbf{k}\cdot \mathbf{p}}    \, ,
\end{multline}
where we have introduced another set of relative coordinates for the centroid phase-space coordinates of the initial wave packets:
\begin{align}
  & & \mathbf{r} &= \mathbf{r}_1 - \mathbf{r}_2 \, , \\
  \mathbf{P_i} &= \mathbf{p}_1+\mathbf{p}_2  \, , &\mathbf{p} &= \frac{1}{2}\left( \mathbf{p}_1 - \mathbf{p}_2 \right)    \, .
\end{align}
The distribution 
\begin{equation}
	\label{eq:jdist}
  J_{\mathbf{P}_i,\mathbf{P}_f} =  \frac{\delta^3}{\pi^{3/2}\hbar^3} e^{-\frac{\delta^2}{\hbar^2}\left(\mathbf{P}_f - \mathbf{P}_i \right)^2}
\end{equation}
describes the overlap of the initial momentum of the two coalescing particles, represented by the centroid momentum $\mathbf{P}_i$, with the final momentum eigenstate characterized by $\mathbf{P}_f$. It is simply a Gaussian with a width $\sqrt{2} \hbar /\delta$, which corresponds to a random superposition of the fluctuations of widths $\hbar/\delta$ around the two initial momenta $\mathbf{p}_1$ and $\mathbf{p}_2$. Thus, the quantum uncertainty in the final momentum of the meson
reflects the quantum uncertainty in the initial quark momenta.

$J_{\mathbf{P}_i,\mathbf{P}_f} $ is properly normalized to one with respect to integration over $\mathbf{P}_f$. For the rest of this section, we will only deal with the probability for
coalescence into any final-state motion of the bound state
\begin{equation}
   \pcal_{kl} = \sum_m \int d^3\mathbf{P}_f \tilde \pcal_{klm,\mathbf{P}_f}  \, .
\end{equation}
We also sum over the magnetic quantum number $m$ as we will not consider the different polarization states of the bound state. From Eq.\ (\ref{eq:pf0}), we obtain
\begin{multline}
\label{eq:pf}
   \pcal_{kl}=  8  e^{-\frac{r^2}{4\delta^2} -4\frac{\delta^2}{\hbar^2} p^2}  
   \int d^3 \Delta \mathbf{x} \, d^3 \Delta \mathbf{k}  \,
   W_{kl} \left( \Delta \mathbf{x} , \Delta \mathbf{k} \right)   \\ \times
   e^{-\frac{\Delta x^2}{4\delta^2} + \frac{1}{2\delta^2} \Delta \mathbf{x}\cdot \mathbf{r}}
   e^{-4\frac{\delta^2}{\hbar^2} \Delta k^2 +8\frac{\delta^2}{\hbar^2} \Delta \mathbf{k}\cdot \mathbf{p}}    \, .
\end{multline}
The result for any particular final-state momentum can be quickly recovered by adding the appropriate factor $J_{\mathbf{P}_i,\mathbf{P}_f}$ to Eq.\ (\ref{eq:pf}).   We note that the coalescence probability $\pcal_{kl}(\mathbf{r},\mathbf{p})$ will simply be a function of the displacement vector of the centroids of two particles in phase space $(\mathbf{r},\mathbf{p})$ and of the final-state quantum numbers $k$ and $l$, as expected.  In an abuse of language, we will, for sake of brevity, sometimes refer to the positions of the centroids as the "positions" and "momenta" of the particles from here on.

Rather than computing the remaining integrals directly using the results from Subsection \ref{sec:3.2}, we once more utilize the expansion into factorized 1-D eigenstates and apply the results for coalescence in one dimension. To this end, we apply the overlap integral Eq.\ (\ref{eq:pf}) to the Wigner distributions in factorized form in Eq.\ (\ref{eq:wignerferep}). After summing over $m$, we obtain a representation of the coalescence probabilities 
\begin{equation}
   \label{eq:pkl}
   \pcal_{kl} =  \sum_{\substack{n_1,n_2,n_3 \\ n'_1,n'_2,n'_3}}  D_{kl}\! \left( \substack{n_1,n_2,n_3 \\ n'_1,n'_2,n'_3} \right)
   \hat P_{n'_1 n_1} \hat P_{n'_2  n_2} \hat P_{n'_3  n_3}\, ,
\end{equation}
where the $\hat P_{n'  n}$ are quasi-probabilities obtained from off-diagonal 1-D Wigner functions $W_{n' n}$ through
the equivalent of Eq.\ (\ref{eq:pf}) in 1-D, i.e. ($i=1,2,3$),
\begin{multline}
   \hat P_{n'  n} = 2  e^{-\frac{r_i^2}{4\delta^2} - \frac{4\delta^2}{\hbar^2} p_i^2}  
   \int d \Delta x_i \, d \Delta k_i  \,
   W_{n' n} \left( \Delta x_i , \Delta k_i \right)   \\ \times
   e^{-\frac{\Delta x_i^2}{4\delta^2} + \frac{1}{2\delta^2} \Delta x_i r_i}
   e^{- 4 \frac{\delta^2}{\hbar^2} \Delta k_i^2 + 8 \frac{\delta^2}{\hbar^2} \Delta k_i p_i}  \, .
\end{multline}
We briefly discuss these 1-D Wigner functions in the following.

\subsection{Computing the Probabilities --- 1-D Case}

Let us recall the generating function Eq.\ (\ref{eq:wgen}). We define a generating function for the quasi-probabilities in analogy with the previous equation as
\begin{multline}
    I(\alpha,\beta;r_i,p_i) =  2 e^{-\frac{r_i^2}{4\delta^2} - \frac{4\delta^2}{\hbar^2} p_i^2}  
   \int d \Delta x_i \, d \Delta k_i  \,
   G(\alpha,\beta;\Delta x_i,\Delta k_i)     \\ \times
   e^{-\frac{\Delta x_i^2}{4\delta^2} + \frac{1}{2\delta^2} \Delta x_i r_i}
   e^{-4\frac{\delta^2}{\hbar^2} \Delta k_i^2 + 8\frac{\delta^2}{\hbar^2} \Delta k_i p_i}  \, .
\end{multline}
The integrals over $\Delta x_i$ and $\Delta k_i$ can then be readily taken. The result is 
\begin{multline}
   \label{eq:I1d}
    I(\alpha,\beta;r_i,p_i) = 2 e^{-\alpha\beta} \frac{2\nu\delta}{1+4\nu^2 \delta^2}  e^{-\frac{r_i^2}{4\delta^2} - 
   \frac{4\delta^2}{\hbar^2} p_i^2}\\
   \times e^{\frac{\left(\frac{r_i}{2\delta} + \sqrt{2}\delta\nu (\alpha+\beta)\right)^2}{1+4\nu^2\delta^2}}
    e^{\frac{\left(4\nu\delta^2\frac{p_i}{\hbar} - \frac{i}{\sqrt{2}} (\alpha-\beta)\right)^2}{1+4\nu^2\delta^2}}  \,.
\end{multline}
We can recover the quasi-probabilities $\hat P_{n'  n}$ for particular states $n'$ and $n$ by applying the derivatives in Eq.\ (\ref{eq:wgen2}) to $I(\alpha,\beta;r_i,p_i)$.

\begin{figure}[t]
\begin{center}
  \includegraphics[width=\columnwidth]{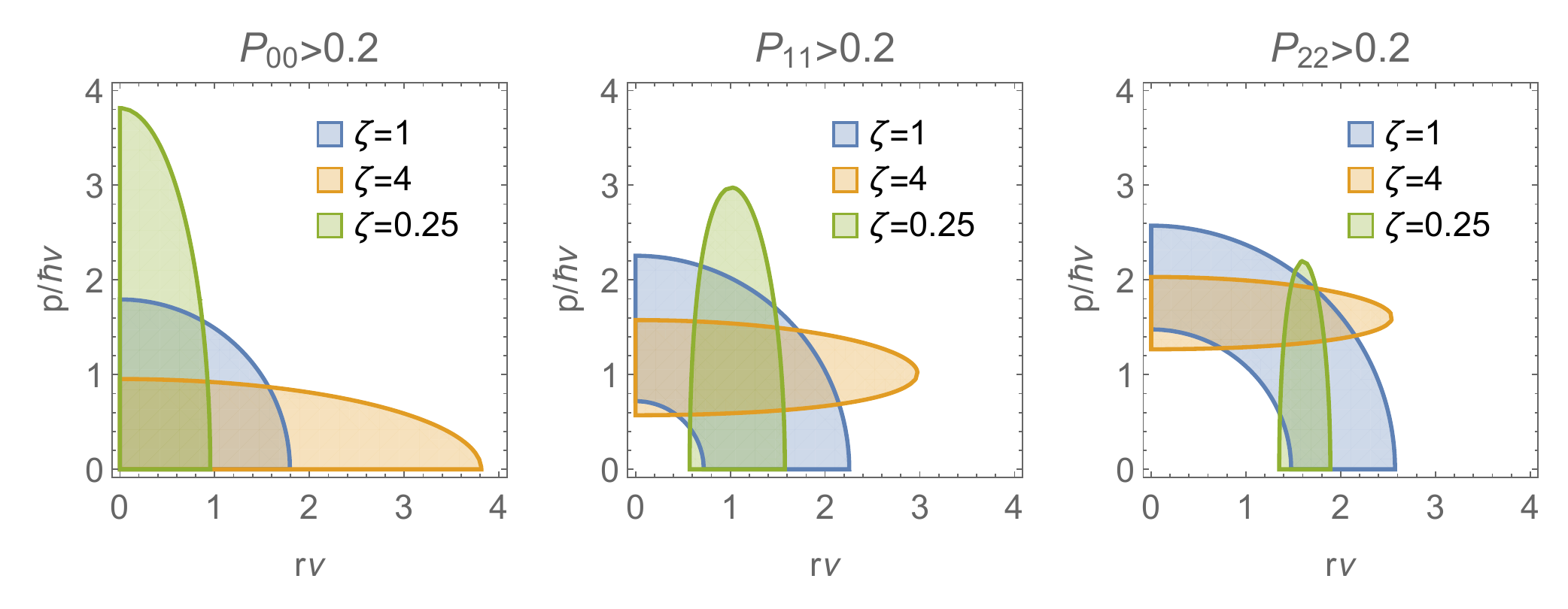}
   \caption{\label{fig:pscale} Areas of probability $\hat P_{nn} \geq 0.2$ for $n=0$ (left panel), $n=1$ (center panel), and $n=2$ (right panel). The symmetric scale ratio $\zeta =1$ is shown as well as $\zeta=4$ and $\zeta=1/4$.}
\end{center}
\end{figure}

The probabilities depend on two scale parameters: the size of the harmonic oscillator potential $1/\nu$ and the size $\delta$ of the incoming wave packets. We will see that the case $\nu^{-1} =2\delta$ plays a special role. It resembles a resonant case in which two wave packets fit neatly into the potential. We define the dimensionless ratio
\begin{equation}
  \zeta = 2\delta \nu  \, .
\end{equation}
We can recast the previous result in terms of $\nu$ and $\zeta$ as
\begin{equation}
   \label{eq:I1d2}
    I =  \frac{2\zeta}{1+\zeta^2}  e^{-\alpha\beta} e^{-\frac{r_i^2 \nu^2}{\zeta^2} - 
   \frac{4 p_i^2 \zeta^2 }{\nu^2\hbar^2}}   
    e^{\frac{\left(\frac{r_i \nu}{\zeta} + \frac{\zeta}{\sqrt{2}} (\alpha+\beta)\right)^2}{1+\zeta^2} +
    \frac{\left(\frac{p_i \zeta^2}{\nu \hbar} - \frac{i}{\sqrt{2}} (\alpha-\beta)\right)^2}{1+\zeta^2}}  \, .
\end{equation}
The pseudo-probabilities for the few lowest orders are
\begin{align}
   \hat P_{00} & = \frac{2\zeta}{1+\zeta^2} e^{-v_i} \\
   \hat P_{01} & = \hat P_{00}  \frac{\sqrt{2}}{(1+\zeta^2) \nu \hbar} \left( i p_i \zeta^2 +r_i \nu^2 \hbar \right) \\
   \hat P_{10} & = \hat P_{01}^*  \\
   \hat P_{11} & = \hat P_{00} \frac{2}{(1+\zeta^2)^2 \nu^2 \hbar^2} \left( p_i^2 \zeta^4 +r_i^2 \nu^4 \hbar^2 \right) 
\end{align}
where
\begin{equation}
  v_i = \frac{r_i^2 \nu^2}{1+\zeta^2} + \frac{p_i^2}{\hbar^2\nu^2 (1+\zeta^2)}  \, .
\end{equation}

Eq.\ (\ref{eq:I1d2}) exhibits an interesting symmetry. The generating function $I$ is invariant under simultaneous replacements 
\begin{equation}
   \zeta \leftrightarrow \frac{1}{\zeta}  \quad , \quad
   r_i \nu \leftrightarrow \frac{p_i}{\nu \hbar} \quad  , \quad (\alpha+\beta) \leftrightarrow i(\alpha-\beta) \, .
\end{equation}
Hence, for the diagonal probabilties $\hat P_{nn}$, for which an equal number of derivatives of $I$ with respect to $\alpha$ and $\beta$ are taken, replacing $\zeta$ by $1/\zeta$ amounts to interchanging the dimensionless coordinate and momentum variables. In other words, $\zeta$  scales the relative contributions that $r_i$ and $p_i$ make to the probability. This is elucidated in Fig.\ \ref{fig:pscale}, which shows the areas of $\hat P_{nn} \geq 0.2$ in the $r_i$-$p_i$-plane for several probabilities. The symmetry around $\zeta=1$ is clearly visible. For $\zeta>1$, i.e. for a bound state size larger than twice the size of the wave packet, recombination favors incoming particles at larger distance and smaller relative momenta, while the opposite is true for $\zeta<0$, when the bound state size is smaller. However, "on average" these cases lead to the same coalescence probabilities. To be more precise, when particle pairs are sampled from a distribution in which $r_i$ and $p_i$ are homogeneously distributed, the probabilities sum to
\begin{equation}
   \int dr_i \int dp_i \, I(\alpha,\beta;r_i,p_i) = 2\pi \hbar e^{\alpha\beta}\, ,
\end{equation}
which is independent of $\zeta$. Thus, while the value of $\zeta$ matters for individual cases, the dependence on $\zeta$ is weaker when applied to distributions of particles, and it disappears in the homogenous case. In the following, we will restrict our discussion to the case $\zeta=1$, in which the algebra simplifies substantially. This is appropriate
for the applications we have in mind, in which the widths of wave packets are usually not well constrained, and for which we ultimately seek statistical answers for large ensembles of particles. However, the more general case can in principle be worked out from the results in this subsection.



In the case $\zeta=1$, the generating function simplifies to
\begin{equation}
    I(\alpha,\beta;r,p) = e^{-\frac{\nu^2 r^2}{2} -\frac{p^2}{2\hbar^2 \nu^2}} 
      e^{\frac{\alpha}{\sqrt{2}}\left(\nu r  - i\frac{p}{\hbar \nu} \right)}
     e^{\frac{\beta}{\sqrt{2}}\left(\nu r  + i\frac{p}{\hbar \nu} \right)}   \, .
\end{equation}
The coalescence probabilities now take particularly simple forms, and they are given by
\begin{equation}
  \hat P_{n' \, n}(r_i,p_i) = \frac{e^{-v_i}}{\sqrt{n! n'!}}  
  \left(\frac{\nu r}{\sqrt{2}} + i \frac{p}{\sqrt{2}\nu\hbar} \right)^n
  \left(\frac{\nu r}{\sqrt{2}} - i \frac{p}{\sqrt{2}\nu\hbar} \right)^{n'}
\end{equation}
where $\zeta$ is set to one in $v_i$.
For $n=n'$ these probabilities have already been derived in Ref.~\cite{Han:2016uhh}. For $n\ne n'$ they are complex, as expected.

\subsection{Computing the Probabilities in the 3-D Case}

\begin{figure}[tb]
\begin{center}
  \includegraphics[width=0.3\columnwidth]{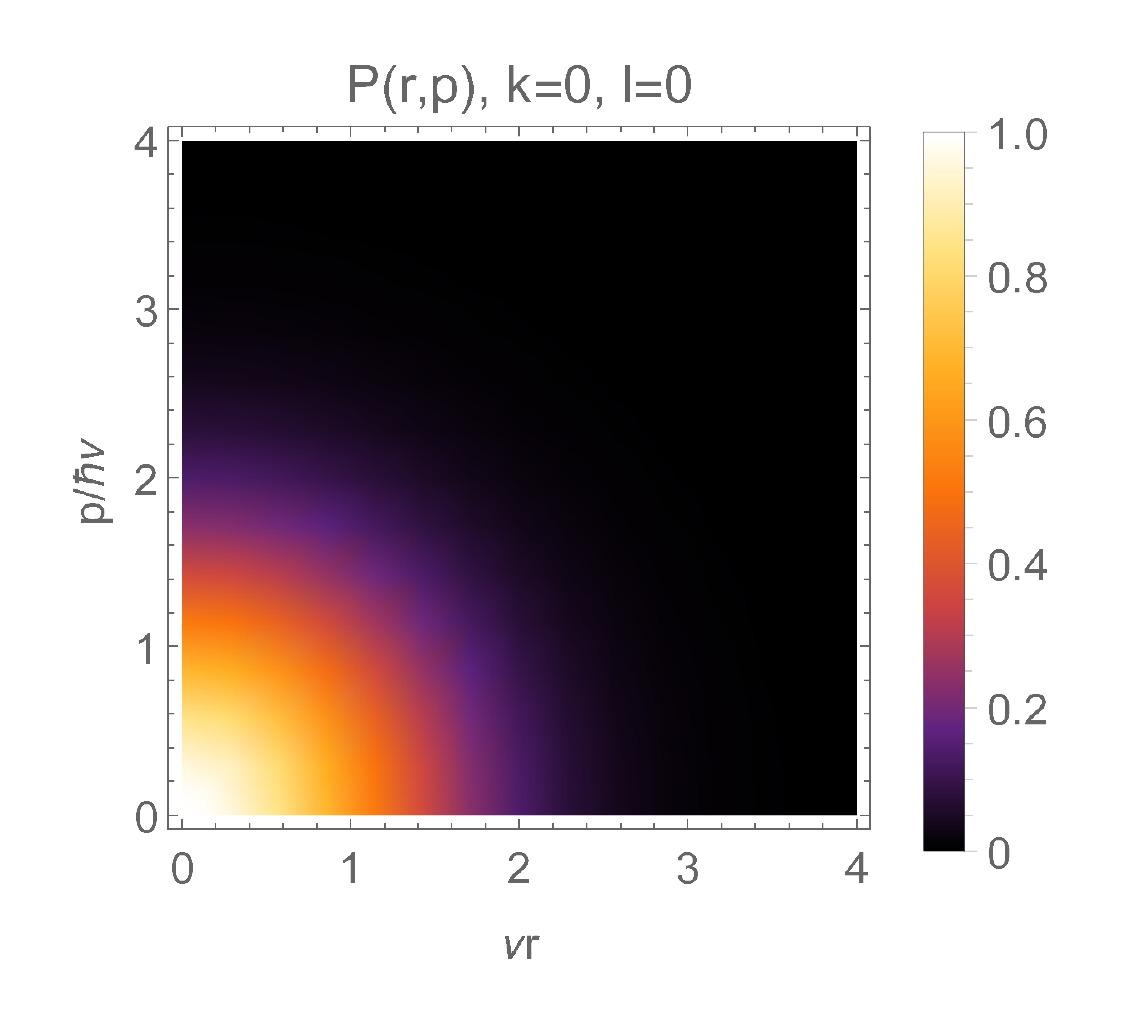}  ~ 
  \includegraphics[width=0.3\columnwidth]{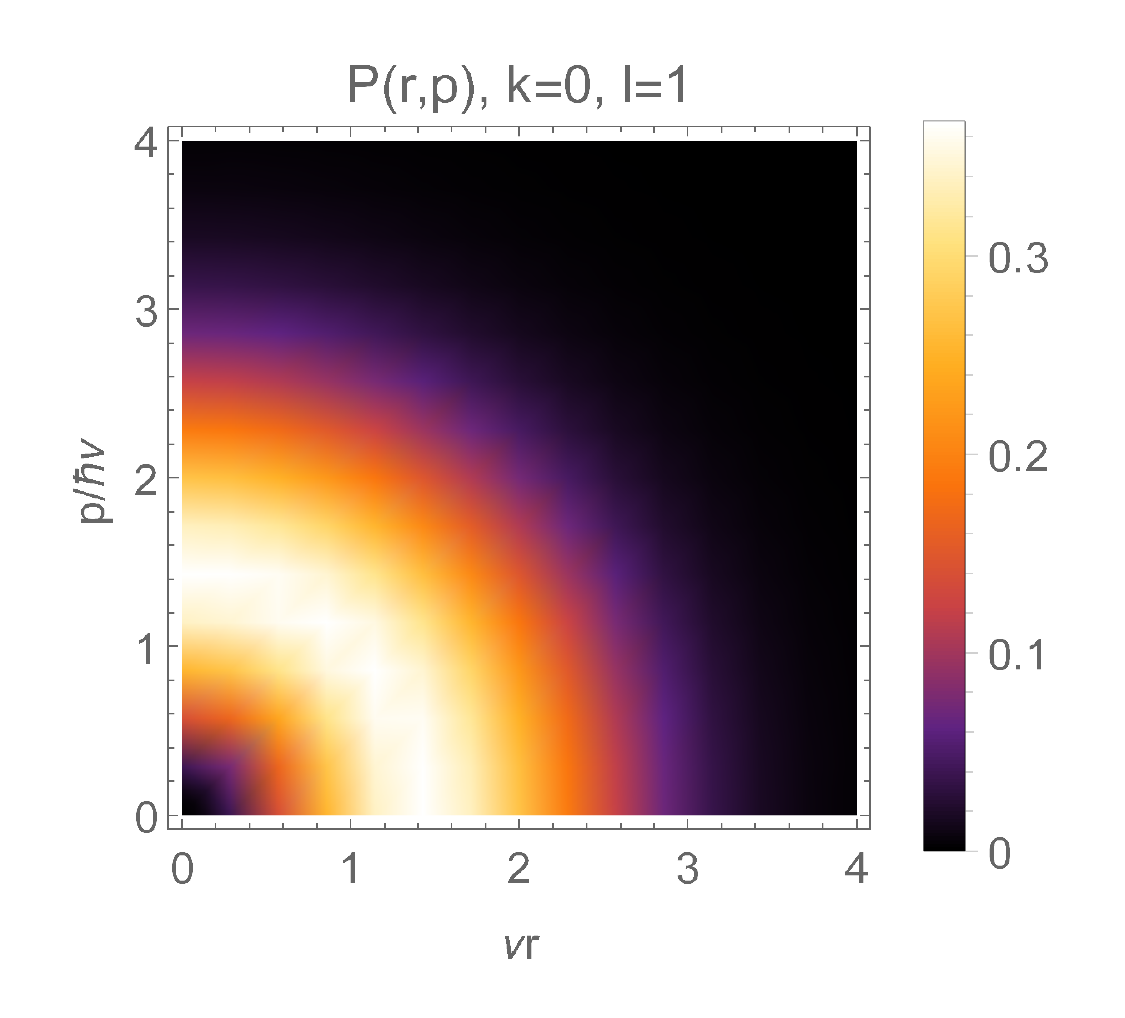} \\
  \includegraphics[width=0.9\textwidth]{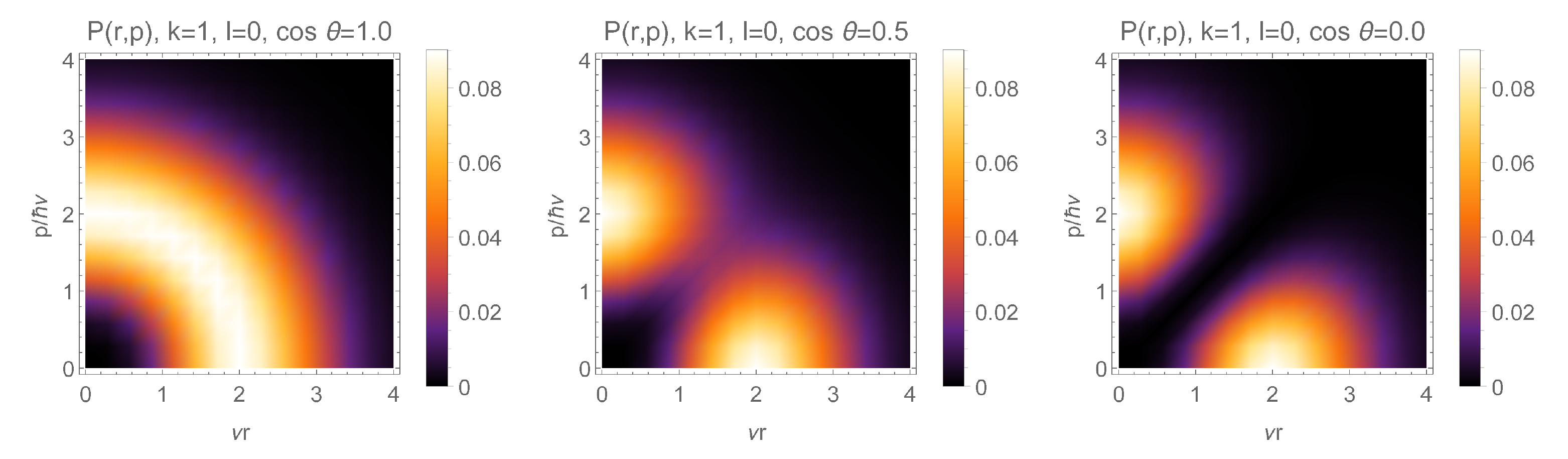} \\
  \includegraphics[width=0.9\textwidth]{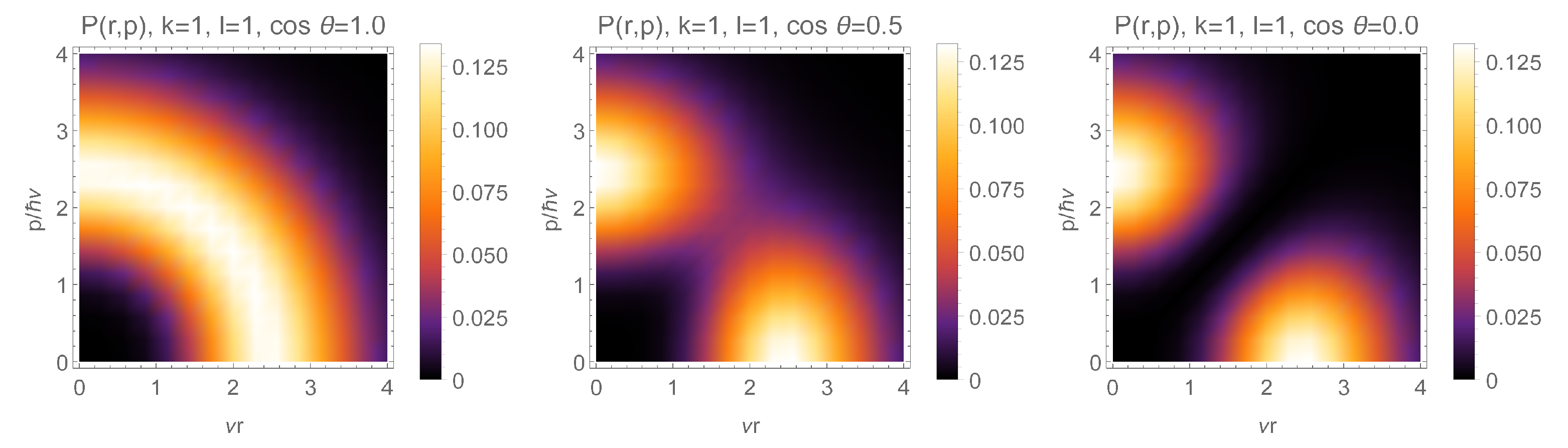}
  \caption{\label{fig:pplots} Coalescence probabilities $P_{kl}$, summed over $m$, for two Gaussian wave packets interacting through an isotropic 3-D harmonic oscillator potential. Plots show probabilities as functions of relative coordinates $r=|\mathbf{r}|$ and $p=|\mathbf{q}|$. Probabilities that depend on the scalar product $\mathbf{r}\cdot \mathbf{p}$ are plotted for several values of the angle $\theta$ given by $\cos\theta = \mathbf{r}\cdot \mathbf{p}/rp$.}
\end{center}
\end{figure}

We can now finally apply Eq.\ (\ref{eq:pkl}) and obtain the coalescence probabilities into angular momentum eigenstates. For the few lowest values of $k$ and $l$, they read
\begin{align}
   \pcal_{00} &= e^{-v} \, ,  \\
   \pcal_{01} &= e^{-v} v \, , \\
   \pcal_{02} &= \frac{1}{2} e^{-v} \left( \frac{2}{3} v^2 +\frac{1}{3} t\right) \, ,  \\
   \pcal_{10} &= \frac{1}{2} e^{-v} \left( \frac{1}{3} v^2 -\frac{1}{3} t\right) \, ,  \\
   \pcal_{03} &= \frac{1}{3!} e^{-v} \left( \frac{2}{5} v^3 +\frac{3}{5} vt\right)\, ,  \\
   \pcal_{11} &= \frac{1}{3!} e^{-v} \left( \frac{3}{5} v^3 -\frac{3}{5} vt\right)  \, .
\end{align}
In the above, we have used the short-hand notations
\begin{align}
  v &= \frac{\nu^2 r^2}{2} + \frac{p^2}{2\hbar^2\nu^2} \, ,   \\
  \label{eq:t}
  t &= \frac{1}{\hbar^2} \left[p^2 r^2 - (\mathbf{p}\cdot \mathbf{r})^2 \right] =  \frac{1}{\hbar^2}  L^2 \, ,
\end{align}
where $\mathbf{ L} = \mathbf{r}  \times\mathbf{p}$ is the relative angular momentum of the initial particles, given by the centroids of their respective wave packets,
and $L=|\mathbf{L}|$.
As in the 1-D case case, $\zeta=1$ leads to results in which the dimensionless distance and momentum differences are symmetric. The case $\zeta\ne 1$ can be treated similarly if needed. In Fig.~\ref{fig:pplots}, we show the coalescence probabilities $P_{kl}$, summed over $m$, for two Gaussian wave packets interacting with an isotropic 3-D harmonic oscillator potential. The plots show probabilities as functions of relative coordinates $r=|\mathbf{r}|$ and $p=|\mathbf{q}|$ as well as their dependence on the scalar product $\mathbf{r}\cdot \mathbf{p}$ for several values of the angle $\theta$ given by $\cos\theta = \mathbf{r}\cdot \mathbf{p}/rp$.

\begin{figure}[t]
\begin{center}
  \includegraphics[width=0.5\columnwidth]{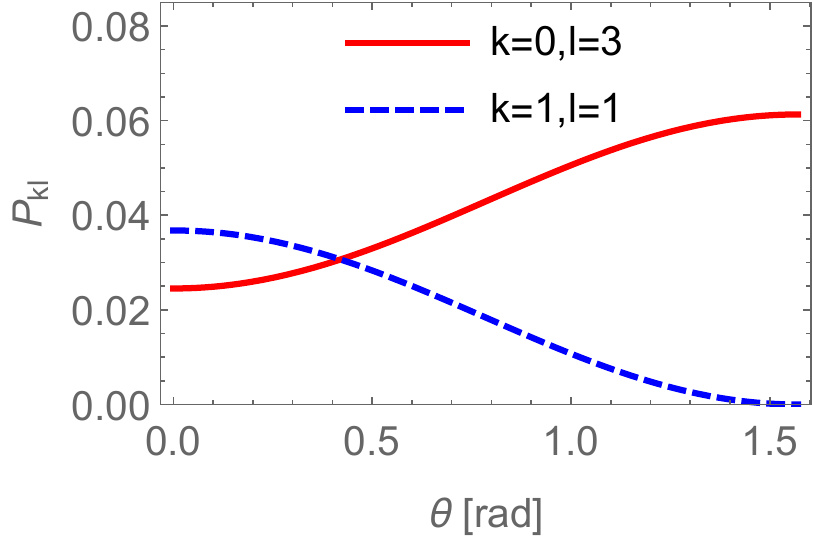}  ~ 
  \caption{\label{fig:theta} Coalescence probabilities $P_{03}$ and $P_{11}$, summed over $m$, as a function of the angle $\theta$ between the relative positions and momenta of the particles, for fixed magnitudes $r=1/\nu$ and $p=\hbar\nu$ of relative position and momentum. The (0,3) and (1,1) states both have energy quantum number $N=3$.
}
\end{center}
\end{figure}

It was first pointed out in Ref.~\cite{Han:2016uhh} that in the 1-D case the probability to coalesce into a state with energy quantum number $n$ is given by the Poisson distribution. It was also shown, using factorization into 1-D eigenstates, that this behavior extends to the 3-D case if $n$ is replaced by the energy quantum number 
$N=n_1+n_2+n_3$. This result directly translates into the probability for coalescence into angular momentum eigenstates with energy $N=2k+l$, since the eigenspace
in the corresponding Hilbert space is the same. Indeed, we confirm this behavior in our calculation. Based on the lowest orders in $N=2k+l$, the probability to coalesce 
into a state with energy $N$ is
\begin{equation}
    \sum_{2k+l=N}  \pcal_{kl} =  e^{-v}\frac{v^N}{N!} \, ,
\end{equation}
As a corollary, unitarity for the coalescence process is obvious. 
For a fixed energy quantum number $N$, the splitting of probabilities between states of different radial and orbital angular momentum quantum numbers is governed by the squared classical angular momentum $t\hbar^2$ of the centroids of the initial wave packets. As expected, larger angular momentum quantum numbers $l$ are favored by larger $t$, defined in Eq.\ (\ref{eq:t}). Fig.\ \ref{fig:theta} compares the coalescence probabilities for two states with $N=3$, $\pcal_{03}$ and $\pcal_{11}$. At fixed values of $r$ and $p$, the probabilities to coalesce into the $k=1$ radially excited states drops with increasing angle $\theta$ between $\mathbf{r}$ and $\mathbf{p}$, reaching a minimum at $\theta=\pi/2$. Conversely, the probability to coalesce into an orbital angular momentum excited state increases with $\theta$. In other words, particles that initially move towards each other, or away from each other ($\theta\approx 0$ or $\pi$) tend to coalesce into lower orbital angular momentum but radially excited states of the appropriate energy, while particles that move orthogonally to their distance vector ($\theta \approx \pm \pi/2$) prefer higher orbital angular momentum bound states. This translation between the initial classical angular momentum $\mathbf{L}$ and the preferred quantum numbers $(k,l)$ is as expected. Thus, we are justified to write the probabilities as $\pcal(v,t)$, indicating their dependence on the squared phase-space distance and squared orbital angular momentum.

\subsection{Applications to Distributions of Coalescing Particles}

Let us briefly turn to the the question of how we can use the results from the previous subsection to compute yields of bound states from the coalescence of particles that are given by classical particle distributions $f(\mathbf{r},\mathbf{p})$. This can be achieved by interpreting, in a semi-classical limit, the distributions to represent the centroid phase-space coordinates of the Gaussian wave packets used in our calculation. Following Eq.\ (\ref{eq:coalp1}), the number of particles with final-state quantum numbers $klm$ and final momentum in an infinitesimal volume around $\mathbf{P}_f$ is
\begin{multline}
 d N_{klm,\mathbf{P}_f}= d^3 \mathbf{P}_f 
   \int d^3 \mathbf{r}_1 d^3 \mathbf{p}_1 f_1(\mathbf{r}_1,\mathbf{p}_1)     \int d^3 \mathbf{r}_2 d^3 \mathbf{p}_2  f_2(\mathbf{r}_2,\mathbf{p}_2) \\
   \times {(2\pi\hbar)}^6 \int d^3 \mathbf{x}_1 d^3 \mathbf{x}_2 d^3 \mathbf{k}_1 d^3 \mathbf{k}_2  W_{\mathbf{P}_f} (\mathbf{K})    \\ \times
   W_{klm}  \left( \Delta \mathbf{x} , \Delta \mathbf{k} \right) 
   W_1(\mathbf{x}_1,\mathbf{k}_1) W_2(\mathbf{x}_2,\mathbf{k}_2)   \, .
  \label{eq:coaldist}
\end{multline}
The Gaussian wave packets $W_i$ are defined as in Eq.\ (\ref{eq:gaussinput}) with centroids $(\mathbf{r}_i,\mathbf{p}_i)$. Using the results from the previous section, we can write the differential yield, summed over spin $m$, as
\begin{equation}
 \frac{dN_{kl,\mathbf{P}_f}}{d^3 \mathbf{P}_f }
  = \int d^3 \mathbf{r}_1 d^3 \mathbf{p}_1 f_1(\mathbf{r}_1,\mathbf{p}_1)     \int d^3 \mathbf{r}_2 d^3 \mathbf{p}_2  f_2(\mathbf{r}_2,\mathbf{p}_2) 
   J_{\mathbf{P}_i ,\mathbf{P}_f} \mathcal{P}_{kl} \, ,
\end{equation}
where $J_{\mathbf{P}_i,\mathbf{P}_f}$ (see Eq.\ (\ref{eq:jdist})) reflects the Gaussian smearing of initial particle momenta. In the semi-classical approximation, $J$ approaches a $\delta$-function $J \to \delta^{(3)}(\mathbf{P}_i -\mathbf{P}_f)$. This formula has often been used in the nuclear physics literature on coalescence, see e.g.\ \ \cite{Sato:1981ez,Baltz:1995tv,Scheibl:1998tk,Mattiello:1996gq,Kahana:1996bw,Chen:2003qj,Chen:2003ava,Zhu:2015voa,Zhao:2018lyf},  and is consistent with our approach in the semi-classical limit of a sharp final momentum for the bound state. 

\section{Summary and Discussion}\label{sec:5}

This work presents three main results. First, Eq.\ (\ref{eq:finalcoeff1}) gives a general expression for the coefficients needed to expand angular momentum eigenstates of the isotropic 3-D harmonic oscillator into factorized eigenstates that utilize products of 1-D eigenstates. A closed form can be given for $k=0$ states. The 1-D harmonic oscillator has been studied exhaustively, and the expansion coefficients pave a path to utilizing 1-D results to derive novel statements for the isotropic 3-D harmonic oscillator. 

Second, we have used this technique to derive Wigner phase-space distributions in terms of their 1-D counterparts, see Eq.\ (\ref{eq:wignerferep}). Explicit expressions are given in Eqs.\ (\ref{eq:wigfinal}) through (\ref{eq:wigfinal2}), which are products of Gaussians in the dimensionless phase-space coordinate $\sqrt{r^2\nu^2 + q^2/(\hbar^2 \nu^2)}$ and polynomials of degree $N$ in $r^2$, $q^2$ and $\mathbf{r}\cdot \mathbf{q}$. The scalar product $\mathbf{r}\cdot \mathbf{q}$ produces a dependence on the relative orientation of the coordinate and momentum vectors. Our results reproduce previous results from the literature.

Finally, we have considered the coalescence of two non-relativistic particles, approximated by Gaussian wave packets, interacting through a harmonic oscillator potential. Coalescence probabilities into angular momentum eigenstates can once more be expressed through the corresponding 1-D coalescence probabilities using the expansion coefficients from Sec.\ \ref{sec:2}. Particularly simple results emerge if the size of the initial wave packets and the width of the harmonic oscillator potential obey a 1:2 ratio ($\zeta = 1$). In that case, the probabilities for coalescence into states with energy quantum number $N$ is given by the Poisson distribution with respect to the energy quantum number $N$ and the squared phase-space distance of the particles, as given by the centroids of the wave packets. Probabilities for different angular momentum states $l$ with the same energy quantum number $N$ are differentiated by the relative angular momentum $L$ of the initial particles. Deviations from $\zeta=1$ lead to a relative rescaling of the relative distance and momentum in the probability, with larger initial wave packets favoring smaller relative distances but allowing for larger relative momentum.

While various applications can be envisioned, the formalism as discussed here readily applies to the coalescence of quark-antiquark pairs into mesons.  Let us briefly sketch out this scenario using mesons as given by the quark model. In spectroscopic notation, these mesons can be classified as $n^{2s+1}L_j$ \cite{ParticleDataGroup:2020ssz}, where $L$ denotes orbital angular momentum by the usual letters $S$, $P$, $D$, $F$, etc., for $l=0,1,2,3, \ldots$, respectively, and $n=k+1$ is the radial quantum number, with "1" referring to the radial ground state. The spin degeneracy of the two quark-system is denoted by $2s+1$, with the eigenvalue $s(s+1)\hbar^2 $ of the squared spin operator $\mathbf{S}^2$. The two spin-$1/2$ quarks can couple to a $s=0$ singlet state or a $s=1$ triplet state. Spin and orbital angular momentum of the quarks further couple to the total angular momentum given by the quantum number $j$, with its 3-component $m_j$, satisfying $|l-s| \le j \le l+s$ as usual. A physical meson state $M=n^{2s+1}L_j$, with meson spin orientation $m_j$, is in general a superposition of pure orbital angular momentum and spin eigenstates
\begin{equation}
  \Psi^{(M)}_{j m_j, l s}   = \sum_{m_j=m+m_s} d_{j m_j;l m ,s m_s}  \Psi_{klm} \chi_{s m_s} \zeta_0  \, ,
\end{equation}
where we have suppressed the dependence of phase-space coordinates in the notation for brevity. $\chi_{s, m_s}$ is the spin wave function, the $d_{j m_j;l m ,s m_s}$ are the appropriate SU(2) Clebsch-Gordan coefficients, and $\zeta_0$ denotes the color singlet SU(3) wave function of the two-quark state. Quarks and antiquarks can be represented similarly by their Gaussian wave packets and superpositions of spin and color eigenstates.

As an example, let us assume a statistical distribution of spins and colors of the initial quarks, which we take to be an up quark $u$ and an anti-down quark $\bar d$. Leaving details of the spin and color algebra to a future discussion \cite{Kordell:2022}, and neglecting the effects of confinement\footnote{Confinement would imply correlations between quark colors and cannot be captured by the assumption of a statistical distribution.}, we simply state the final probabilities for coalescence of these quarks if their relative squared phase-space distance is $v$ and squared relative angular momentum is $t$. For the lowest mass $u\bar d$-meson states, pions and $\rho$-mesons as well as their heavier relatives, the probabilities are
\begin{align}
    \pi^+ \quad 1^{1}S_0: & \quad\frac{1}{4\times 9} \pcal_{00}(v,t)   \, , \\
    \rho^+   \quad 1^{3}S_1: & \quad\frac{3}{4\times 9} \pcal_{00}(v,t)   \,  , \\
    b_1^+(1235)   \quad 1^{1}P_1: & \quad\frac{1}{4\times 9} \pcal_{01}(v,t) \, , \\
    a_0^+(1450) \quad  1^{3}P_0: &\quad \frac{3}{4\times 9\times 9} \pcal_{01}(v,t)  \, , \\
    a_1^+(1260)   \quad1^{3}P_1: & \quad\frac{3\times 3}{4\times 9\times 9} \pcal_{01}(v,t)  \, , \\
   a_2^+(1320)   \quad  1^{3}P_2: & \quad\frac{3\times 5}{4\times 9\times 9} \pcal_{01}(v,t) \, , \\
   \pi^+(1300) \quad 2^{1}S_0: & \quad\frac{1}{4\times 9} \pcal_{10}(v,t)   \, , \\
   \rho^+(1450)   \quad 2^{3}S_1: & \quad\frac{3}{4\times 9} \pcal_{10}(v,t)   \,  ,
\end{align}   
The widths $1/\nu$ of the wave functions can be inferred in many cases from the measured charge radii of the mesons \cite{Han:2016uhh,Kordell:2022}.
We expect to report on further applications to the coalescence of quarks soon elsewhere \cite{Kordell:2022}.

{\bf Acknowledgements.}

We are grateful to S.\ Cho for helpful comments. This work was support by the U.S.\ National Science Foundation under awards 1812431 and 2111568, the U.S.\ Department of Energy under Award No.\ DE-SC0015266 and the Welch Foundation under Grant No. A-1358.




\bibliography{3DHO_Theory}

\end{document}